\documentclass[aps, prl, reprint, superscriptaddress]{revtex4-1}
\usepackage{bm} 
\usepackage{dcolumn} 
\usepackage{graphicx} 
\usepackage{amsmath} 
\usepackage{amssymb} 
\usepackage{amsfonts} 
\usepackage{placeins}

\begin{document}
\title{Theory of anisotropic plasmons}

\author{Seongjin Ahn}
\affiliation{Condensed Matter Theory Center and Joint Quantum Institute, Department of Physics, University of Maryland, College Park, Maryland 20742-4111, USA}
\author{S. Das Sarma}
\affiliation{Condensed Matter Theory Center and Joint Quantum Institute, Department of Physics, University of Maryland, College Park, Maryland 20742-4111, USA}

\date{\today}

\begin{abstract}
We develop the complete theory for the collective plasmon modes of an interacting electron system in the presence of explicit mass (or velocity) anisotropy in the corresponding non-interacting situation, with the effective Fermi velocity being different along different axes.  Such effective mass anisotropy is common in solid state materials (e.g., silicon or germanium), where the Fermi surface is often not spherical. We find that the plasmon dispersion itself develops significant anisotropy in such systems, and the commonly used isotropic approximation of using a density of states or optical effective mass does not work for the anisotropic system.  We predict a qualitatively new phenomenon in anisotropic systems with no corresponding isotropic analog,  where the plasmon mode along one direction  decays into electron-hole pairs through Landau damping while the mode remains undamped and stable along a different direction.

\end{abstract}

\maketitle

Plasmons are collective modes of interacting electron systems with Coulomb coupling.  They are the quantum versions of plasma waves in charged plasmas in the way phonons are quantum versions of sound waves in matter.  The long-range Coulomb coupling between the electrons produces a self-sustaining oscillation if the charge density is disturbed, and quantum mechanically, this leads to an elementary collective excitation, the ``plasmon’'. Plasmons have been extensively studied in three dimensional (3D), two-dimensional (2D), and one-dimensional (1D) electron materials for almost 70 years, both theoretically and experimentally, and plasmons are of technological interest in the context of creating intense and controllable confined electric fields in the applied subject of ``plasmonics'' \cite{Maier2007}. In fact, the modern many-body theory, involving the study of electron-electron interactions in solids, started as a subject in the early 1950s with the focus on the plasmon modes in the Bohm-Pines collective coordinates approach \cite{Pines1952,Bohm1953}.   It is probably not an exaggeration to say that plasmons are among the most-studied elementary excitations in all of condensed matter physics \cite{Pines1966,Platzman1973,Abrikosov1988,Quinn2009,Abrikosov2012,Fetter2012,Mahan2013,Pines2018}. The Physical Review journal series has more than 5000 publications with the word ‘plasmons’ in the title or abstract over the last 60 years.  Essentially all the theoretical techniques in the arsenal of solid state (or condensed matter) physics have been used to study plasmons, in fact, most of these techniques were developed first in the study of plasmons in interacting electron liquids, including collective coordinates \cite{Pines1952,Bohm1953}, dielectric function \cite{Lindhard1954, Nozieres1958,Nozieres1958a,Nozieres1958b}, self-consistent field \cite{Ehrenreich1959, Singwi1968}, and diagrammatic field theories \cite{Sawada1957, DuBois1959a, DuBois1959}.  

Given this extensive a background literature spanning more than 60 years, it would appear that all the classic problems in the theory of plasmons in condensed matter physics have already been studied, and new plasmon research is likely to focus only on the collective properties of newly made materials, which, by definition, could not have been studied before, e.g., 2D graphene \cite{Hwang2007, Sarma2009, Yu2018, Brey2020}. This is, however, untrue as the plasmon properties in an anisotropic electron gas (EG) have never been studied before in any depth.  In particular, plasmon properties of an anisotropic 2D (3D) (ellipsoidal) EG, with an elliptic rather than a circular (spherical) isotropic Fermi surface with effective masses $m_x$ and $m_y$ unequal, have been approximated by the corresponding isotropic 2DEG plasmon dispersion \cite{Stern1967}, with the effective mass, $m$, approximated as isotropic and chosen simply as some kind of an average of $m_x$ and $m_y$, with both the density-of-states average $(m_x m_y)^{\frac{1}{2}}$ and the optical average $2m_xm_y/(m_x + m_y)$ being used in the isotropic approximation without any real justification.  This is surprising since many electronic materials, even very simple ones, e.g., Si, Ge, manifest effective mass anisotropy.  In the current work, we address this open question in the plasmon literature, and obtain the theoretical plasmon properties in 2DEG and 3DEG anisotropic systems taking into account the explicit effective mass anisotropy within the random phase approximation (RPA), which is the standard approximation for studying plasmon properties in interacting electron systems. Our work thus puts the theory of plasmons for anisotropic systems on the same footing as that for isotropic systems, correcting an important oversight in the literature.  We find that the isotropic approximation using an averaged effective mass fails quantitatively for the plasmon dispersion of the anisotropic system, and more importantly, the isotropic approximation misses a qualitative plasmon damping effect intrinsic to the anisotropic system, where plasmon damping itself could be anisotropic, with the critical wave vector for plasmon damping threshold through particle-hole pair creation (the so-called Landau damping) depending sensitively on the anisotropy, differing significantly along the different directions on the Fermi surface.

The plasmon is a collective mode of the interacting electron system, and as such it is a pole for the reducible polarizability function. This is equivalent to a zero in the dynamical dielectric function of the system, which is defined by:

\begin{equation}
    \varepsilon(\bm q,\omega)=1-v_\mathrm{c}(\bm q)\Pi(\bm q,\omega)
    \label{eq:diel}
\end{equation}
where $\Pi(q,\omega)$ is the electron irreducible polarizability function (the so-called electron-hole bubble diagram) and $v_\mathrm{c}(q)$ is the Fourier transform of the long-range Coulomb interaction (falling off as 1/distance independent of dimensionality) in the appropriate dimension going as $1/q$ ($1/q^2$) in 2D (3D). RPA, which is the standard approximation for obtaining plasmon properties, involves using the noninteracting irreducible polarizability, $\Pi=\Pi_0$, in Eq.~(\ref{eq:diel}). RPA is an extensively used theory for plasmons, which is exact at long wave length (small $q$). In order to obtain the anisotropic plasmon properties, we must first calculate the irreducible electron polarizability for a system with anisotropic effective mass. Since the polarizability is in general complex for arbitrary wave vector $q$ and frequency $\omega$, the complete solution of Eq.~(\ref{eq:diel}) requires both real and imaginary parts of the dielectric function to vanish at the plasmon frequency (which is a function of the wave vector $q$). This can only happen as long as the plasmon cannot decay by creating electron-hole pairs, i.e., when the plasmon dispersion is completely outside the electron-hole continuum single particle excitation regime of the Fermi surface. Otherwise, the plasmon is Landau-damped through the spontaneous emission of electron-hole pairs, and it is no longer a well-defined collective mode. This implies a real solution of Eq. (1) providing the plasmon dispersion, but the imaginary part of $\Pi(q,\omega)$ is finite, implying Landau damping. An important prediction of our theory is that, in addition to the plasmon being anisotropic by virtue of the mass anisotropy, it also manifests anisotropic damping, and the onset of Landau damping depends explicitly on the direction of the wave vector. Thus, an undamped plasmon could become damped just by virtue of changing its direction of propagation without any change in the magnitude of the wave vector.

We consider first the standard parabolic 2DEG and 3DEG systems with anisotropic band dispersion ($m_\mathrm{H}$ and $m_\mathrm{L}$ being the heavy and light masses with $m_\mathrm{H}>m_\mathrm{L}$) along different axes (with the 3D system being chosen with a layered material in mind so that the mass along the $x$-direction being higher than that in the $y$-$z$ plane):

\begin{equation}
    \varepsilon_\mathrm{2D}(\bm k)=\frac{k_x^2}{2m_\mathrm{H}}+\frac{k_y^2}{2m_\mathrm{L}},
\end{equation}
\begin{equation}
    \varepsilon_\mathrm{3D}(\bm k)=\frac{k_x^2}{2m_\mathrm{H}}+\frac{k_y^2}{2m_\mathrm{L}}+\frac{k_z^2}{2m_\mathrm{L}}.
\end{equation}
We have explicitly calculated the irreducible noninteracting anisotropic polarizability for 2DEG and 3DEG. Then, we solve Eq.~(\ref{eq:diel}) to obtain the 2D and 3D anisotropic plasmon dispersion and plasmon damping as a function of $q_x$ and $q_y$ for a given $q=(q_x^2 + q_y^2)^{1/2}$ in 2D and $(q_x^2 + q_y^2+q_z^2)^{1/2}$ in 3D, using anisotropic effective masses $m_x$ and $m_y$.  Before we present our full numerical results, we provide the analytical long wavelength plasmon dispersion results for 2D [Eq.~(\ref{eq:2D_plasma_frequency}) below] and 3D [Eq.~(\ref{eq:3D_plasma_frequency}) ]:

\begin{equation}
    \omega^{2\mathrm{D}}_\mathrm{p}(\bm q\rightarrow0)=
    \sqrt{\frac{2\pi ne^2q}{m_\mathrm{DOS}}}
    \left| \bm M^{-1/2}\hat{\bm n} \right|,
    \label{eq:2D_plasma_frequency}
\end{equation}
\begin{equation}
    \omega^{3\mathrm{D}}_\mathrm{p}(\bm q\rightarrow0)=
    \sqrt{\frac{4 \pi n e^2}{m_\mathrm{DOS}}}
    \left| \bm M^{-1/2}\hat{\bm n} \right|.
    \label{eq:3D_plasma_frequency}
\end{equation}
Here $\hat{\bm n}=\bm q/|\bm q|$ is the plasma propagation direction, $M_{ij}=m_i/m_\mathrm{DOS}\delta_{ij}$ is the mass matrix and $m_\mathrm{DOS}$ is the so-called density of states mass defined by: $m_\mathrm{DOS}=(m_\mathrm{H} m_\mathrm{L})^{1/2}$ in 2D and $(m_\mathrm{H} m_\mathrm{L}^2)^{1/3}$ in 3D. Note that typically all anisotropic systems are uncritically treated assuming an isotropic density of state mass $m$, whereas in our work we keep the full anisotropy explicitly in the theory. 
\begin{figure}[!htb]
    \centering
    \includegraphics[width=0.9\linewidth]{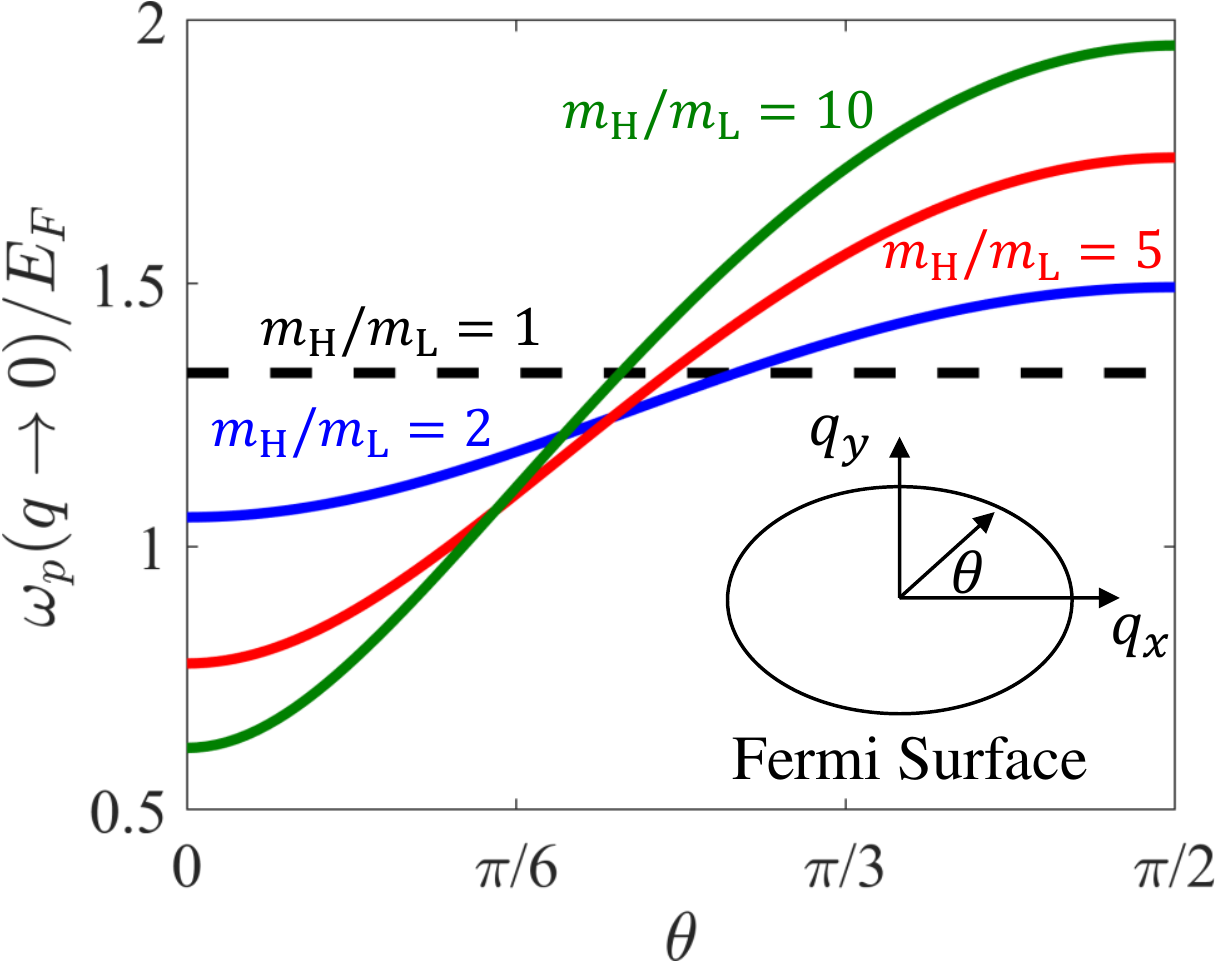}
    \caption{A plot of 3D plasma frequency in the long wavelength limit as a function of $\theta$ for different mass ratios $m_\mathrm{H}/m_\mathrm{L}=1,2,5,10$ and $r_s=2.0$. The angle $\theta$ is defined with respect to the high-mass axis. Here we set $q_z=0$ for simplicity, without loss of generality.
      }
    \label{fig:longwavelength_plasma_frequency}
\end{figure}
\begin{figure*}[!htb]
    \centering
    \includegraphics[width=1.0\linewidth]{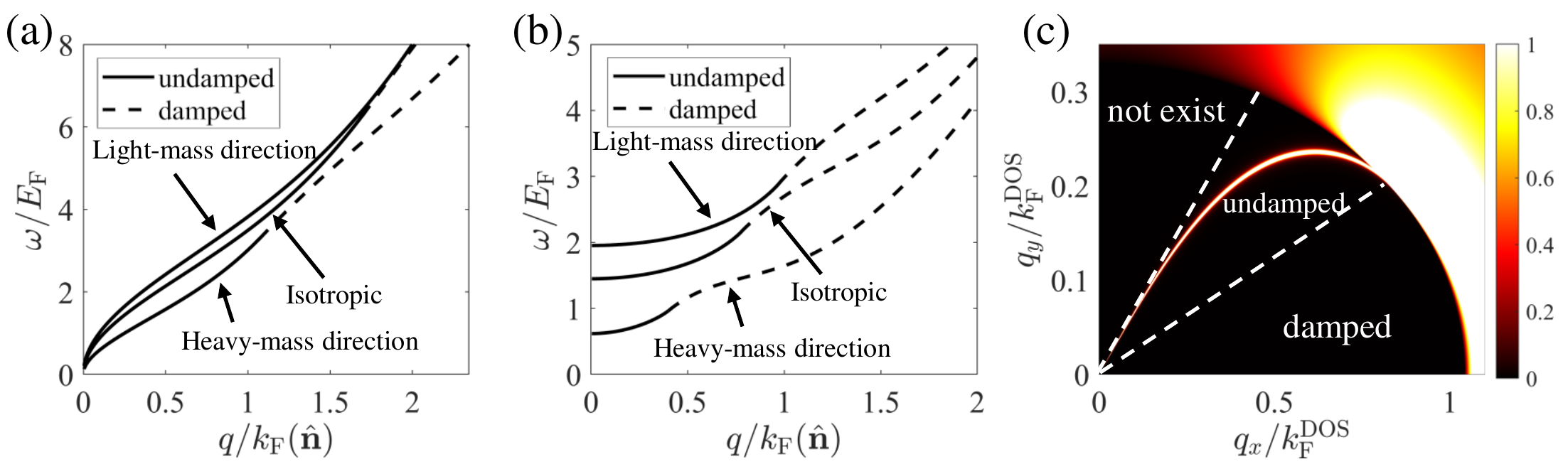}
    \caption{(a) 2D and (b) 3D plasmon dispersions in the high- and low-mass directions along with the isotropic dispersion using the density of state mass. (c) 3D energy-loss function (i.e., $\left|\mathrm{Im}\left[\varepsilon(q_x,q_y,\omega)^{-1} \right]\right|$) for a fixed $\omega=1.2E_\mathrm{F}$ on the $q_z=0$ cross-section of the momentum space. The two dashed lines divide the momentum space with $\mathrm{Im}\varepsilon =0$ into three regions, in each of which plasmons are damped, undamped and absent. 
    Here $k_\mathrm{F}(\theta)$ is the magnitude of the Fermi wave vector along the plasma propagating direction $\hat{\bm n}$, and $k_\mathrm{F}^\mathrm{DOS}=\sqrt{2m_\mathrm{DOS}E_\mathrm{F}}$. We set $r_s=2.0$ and $m_\mathrm{H}/m_\mathrm{L}=10$ for the calculations.
      }
    \label{fig:plasmon_dispersion}
\end{figure*}

The long wavelength exact plasmon dispersions in Eqs.~(\ref{eq:2D_plasma_frequency}) and (\ref{eq:3D_plasma_frequency}) go explicitly as $n^{1/2}$ in density, showing the well-known isotropic plasma frequencies $(2\pi ne^2q/m)^{1/2}$ in 2D and $(4 \pi n e^2/m)^{1/2}$ in 3D, but the anisotropic dispersions have explicit and nontrivial anisotropy dependence through $\hat{\bm n}$ and $m_\mathrm{H}$, $m_\mathrm{L}$, which are missed in the isotropic approximation assuming a single density of states effective mass $m_\mathrm{DOS}$. As is well-known, the strict long wavelength ($q=0$) plasmon has a gap in 3D and vanishes in 2D by virtue of the fact that the 3D Coulomb potential is truly long-ranged---this is obvious in Eqs. (5) and (6). In order to clearly bring out the nontrivial plasmon anisotropy, we plot in Fig.~\ref{fig:longwavelength_plasma_frequency} the 3D $q=0$ plasma frequency as a function of the mode direction $\theta$ for different values of the mass anisotropy parameter $m_\mathrm{H}/m_\mathrm{L}$ for $r_s=2.0$, also showing by the horizontal dashed line the corresponding isotropic result using the density of states mass as employed in all theoretical works so far.
Here the dimensionless quantity $r_s$ is the Wigner-Seitz radius defined through $n^{-1}=\pi (r_s a_B)^2$ in 2D and $n^{-1}= (4\pi/3) (r_s a_B)^3$ in 3D, where $a_B=\hbar^2/(m_\mathrm{DOS}e^2)$ is the Bohr radius and $n$ is the relevant electron density. In the following we set $q_z=0$ for 3D for simplicity, without loss of generality. It is manifestly clear that the isotropic approximation is very poor, even qualitatively, in defining the plasma frequency at $q=0$ for the anisotropic system. The actual plasma frequency could be either below or above the isotropic plasma frequency manifesting nontrivial angle dependence.

In Figs.~\ref{fig:plasmon_dispersion} (a) and (b), we depict our numerically calculated anisotropic plasmon dispersions along the high- and low-mass directions in both (a) 2D and (b) 3D by solving Eq.~(\ref{eq:diel}) directly for $\theta =0$ and $\pi/2$, comparing them with the corresponding isotropic dispersion using the density of states mass. Here $k_\mathrm{F}(\hat{\bm  n})=k_\mathrm{F}^\mathrm{DOS}\left| M^{-1/2}\hat{\bm n} \right|^{-1}$ is the magnitude of the Fermi wave vector along the plasma propagating direction $\hat{\bm n}$, where $k_\mathrm{F}^\mathrm{DOS}=\sqrt{2m_\mathrm{DOS}E_\mathrm{F}}$.
Clearly, the isotropic approximation fails not only quantitatively, but also qualitatively. In particular, the plasmon along the heavy-mass direction gets damped at a much smaller wave number than in the light-mass direction by developing a finite imaginary part in the solution corresponding to Eq. (1), i.e., no pure complex poles with both real and imaginary parts vanishing exist for Eq. (1) indicating Landau damping of the plasmon mode. Thus, the anisotropic plasmon not only has very different frequencies at the same wave number along the heavy- and light-mass directions, it also suffers Landau damping very differently along the two directions. It is also worth noting that for a fixed frequency 3D plasmon modes are allowed to exist only along certain directions because the long wavelength plasma frequency is anisotropic. This is clearly seen from Fig.~\ref{fig:plasmon_dispersion} (c), where we plot the calculated energy-loss function for a fixed $\omega=1.2E_\mathrm{F}$ on the $q_z=0$ cross section of the momentum space. Using Eq.~(\ref{eq:3D_plasma_frequency}), we can easily obtain the range of frequencies where plasmons exist along only certain directions:
\begin{equation}
    \sqrt{\frac{4\pi ne^2}{m_\mathrm{H}}}<\omega<\sqrt{\frac{4\pi ne^2}{m_\mathrm{L}}}.
\end{equation}

\begin{figure}[!htb]
    \centering
    \includegraphics[width=1.0\linewidth]{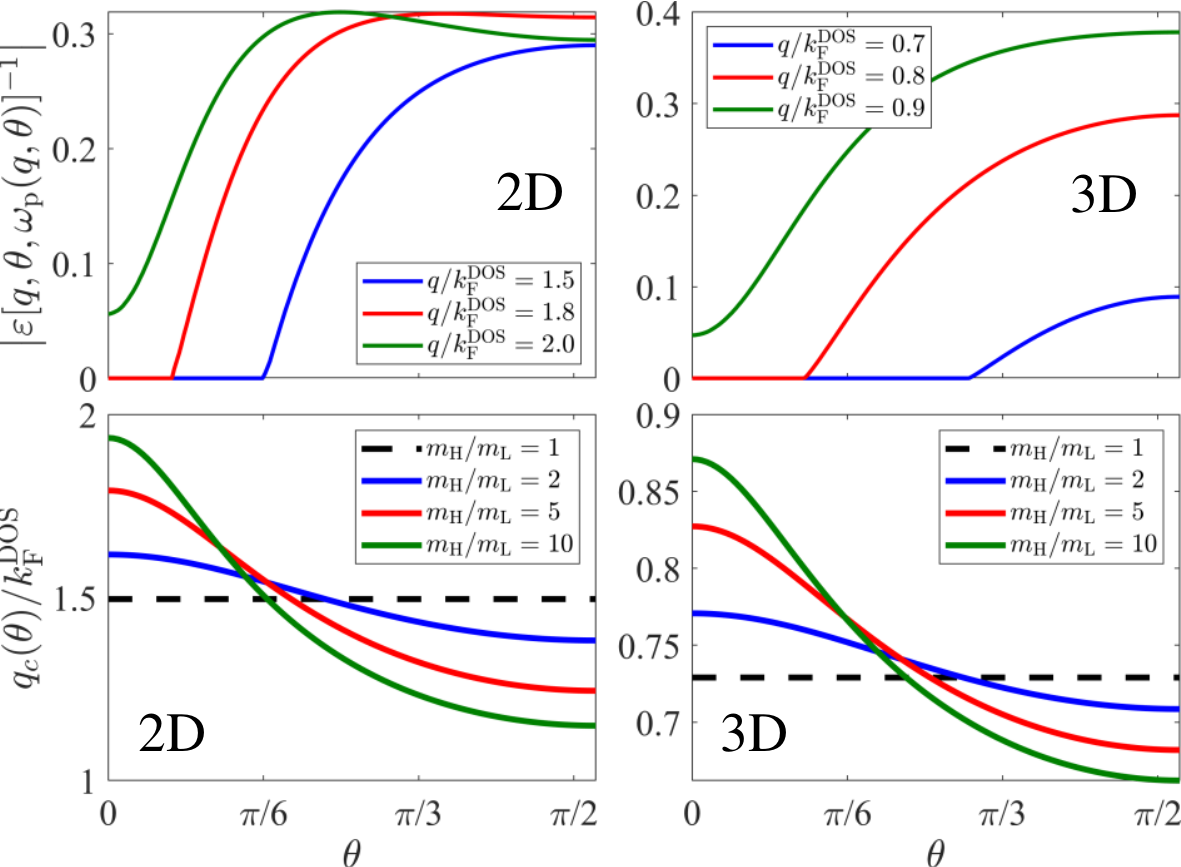}
    \caption{The upper two figures show calculated energy loss functions at the plasma frequency as a function of the direction $\theta$ for $q/k_\mathrm{F}^\mathrm{DOS}=(1.4,1.8,2.0)$ for 2D and $(0.7,0.8,0.9)$ for 3D. The lower two figures show the critical wave vector $k_c$ as a function of the direction $\theta$ for $m_\mathrm{H}/m_\mathrm{L}=1,2,5,10$. For the calculations, we use $r_s=2.0$ and set $q_z=0$ for 3D for simplicity without loss of generality.
      }
    \label{fig:loss_critical}
\end{figure}

\begin{figure}[!htb]
    \centering
    \includegraphics[width=0.84\linewidth]{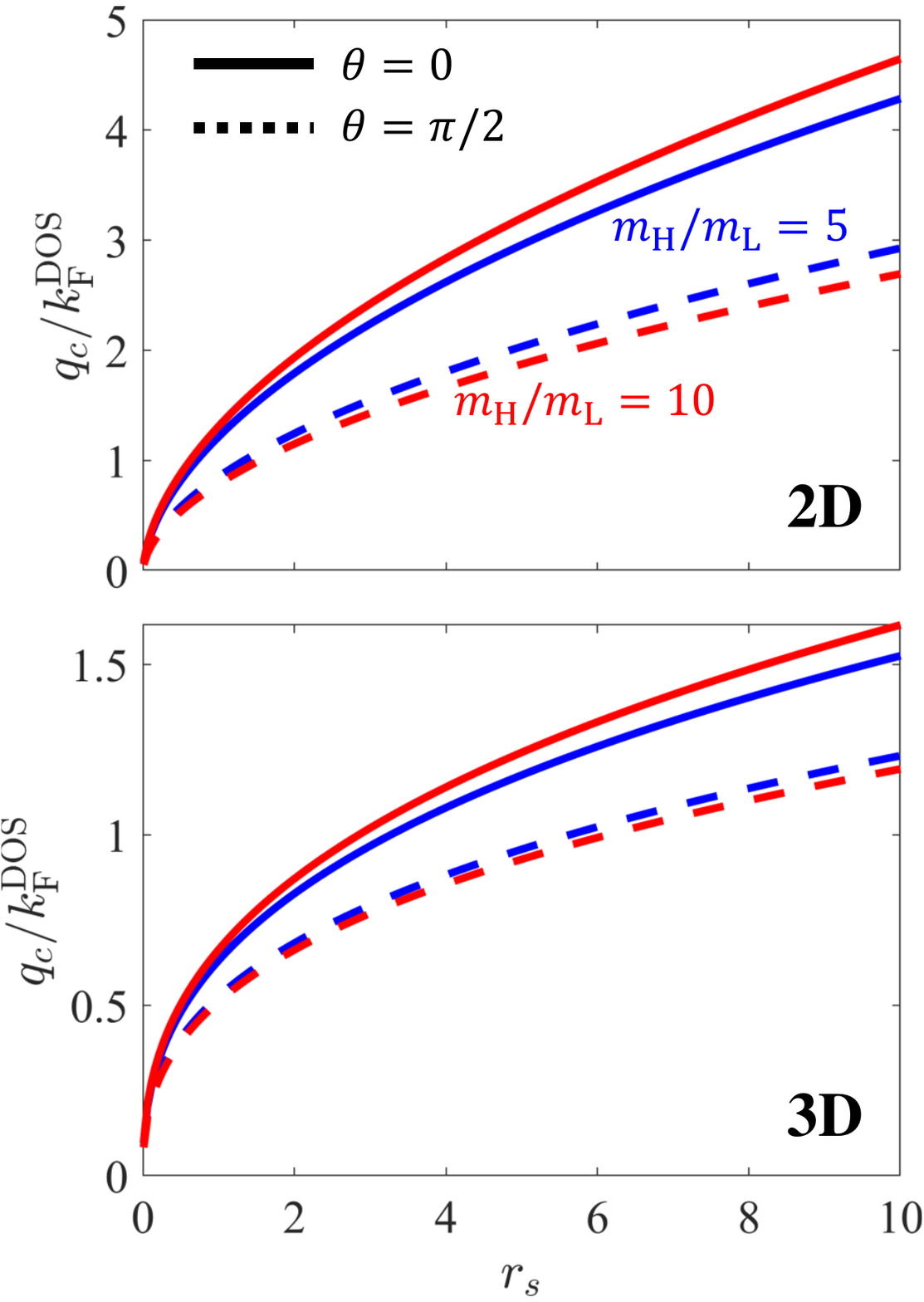}
    \caption{Calculated critical wave vectors at $\theta=0$ (solid) and $\theta=\pi/2$ (dashed) with $m_\mathrm{H}/m_\mathrm{L}=5$ and $10$ as a function of $r_s$.
      }
    \label{fig:kc_theta}
\end{figure}

The dramatic behavior in the Landau damping of anisotropic plasmons is highlighted in Fig.~\ref{fig:loss_critical}. The upper two figures show the calculated loss function, i.e., $\mathrm{Im} (1/\varepsilon)$, at the plasmon frequency as a function of the direction $\theta$ for a few fixed values of the wave number, clearly demonstrating that the Landau damping onset depends crucially on $\theta$ even if the wave number is fixed. By contrast, in the isotropic system, there is just a critical wave number, $k_c$, above which the plasmon is Landau damped because the plasmon dispersion enters the particle-hole continuum at $k_c$. For anisotropic plasmons, however, there is no strict $k_c$ since the Landau damping now depends also on $\theta$, the direction of mode propagation in addition to depending on the magnitude of the wave number. The critical wave number $k_c$ now depends explicitly, not only just on $m_\mathrm{H}/m_\mathrm{L}$, but also on $\theta$ as shown in the lower two figures in Fig.~\ref{fig:loss_critical}. 

It is easy to see from the lower two figures of Fig.~\ref{fig:loss_critical} that that plasmons only with the wave vector $q$ within the range of $q_c(\theta=\pi/2)<q<q_c(\theta=0)$ exhibit anisotropic plasmon damping. Outside this range, plasmons are either damped or undamped along all directions. To analyze this, we begin with the equations determining the critical wave vector $q_c$ given by
\begin{equation}
    \frac{\widetilde{q}_c^2}{\sqrt{2}r_s} + \frac{\widetilde{q}_c^3}{4r_s^2}=
    \frac{1}{\left| \bm M^{-1/2}\hat{\bm n} \right|}
\end{equation}
for 2D and
\begin{align}
     \left(1 + \frac{\left| \bm  M^{-1/2}\hat{\bm n} \right|\widetilde{q}_c}{2}\right)
    \log\left({1+\frac{2}{\left| \bm  M^{-1/2}\hat{\bm n} \right|\widetilde{q}_c}}\right) \nonumber \\ 
    =2\left(\frac{q_c}{k_\mathrm{TF}}\right)^2+1
\end{align}
for 3D, where $k_\mathrm{TF}=2/\sqrt{ \pi(4/9\pi)^{1/3} r_s }a_B$ is the Thomas-Fermi screening wave vector, and $\widetilde{q}=q/k_\mathrm{F}^\mathrm{DOS}$. For $\theta=0$ and $\pi/2$, $\left|M^{-1/2}\hat{\bm n}\right|=(m_\mathrm{L}/m_\mathrm{H})^{(d-1)/2d}$ and $(m_\mathrm{H}/m_\mathrm{L})^{1/2d}$, respectively, where $d$ is the dimension. Using this and taking the high mass ratio limit, we can obtain approximate solutions for $q_c$:
\begin{equation}
\begin{aligned}
    \widetilde{q}_c\left(\theta=0\right)&=2^{2/3}r_s^{2/3}\left(\frac{m_\mathrm{H}}{m_\mathrm{L}}\right)^{1/12},\\
    \widetilde{q}_c\left(\theta=\frac{\pi}{2}\right)&=2^{1/4}r_s^{1/2}\left(\frac{m_\mathrm{L}}{m_\mathrm{H}}\right)^{1/8},
\end{aligned}
\end{equation}
for 2D and
\begin{equation}
\begin{aligned}
    \widetilde{q}_c\left(\theta=0\right)&
    =\frac{(\frac{2}{3})^{1/3}r_s^{1/2}}{\pi^{2/3}}W\left[\frac{(144\pi^4)^{1/3}}{e^2r_s}\left(\frac{m_H}{m_L}\right)^{3/2} \right],\\
    \widetilde{q}_c\left(\theta=\frac{\pi}{2}\right)&
    = \left(\frac{32}{9\pi^4}\right)^{1/9}r_s^{1/3}\left(\frac{m_\mathrm{L}}{m_\mathrm{H}}\right)^{1/18},
\end{aligned}
\end{equation}
for 3D, where $W(x)$ is the Lambert $W$-function defined as the inverse function of $f(x)=x e^x$.
From the obtained asymptotic forms, it is easy to see that the range of wavevector where plasmons are anisotropically damped, i.e., $\widetilde{q}_c\left(\theta=\pi/2\right)<\widetilde{q}<\widetilde{q}_c\left(\theta=0\right)$, becomes wider with increasing $r_s$ and the mass ratio $m_\mathrm{H}/m_\mathrm{L}$. Even though these results are obtained with a very large mass ratio assumed, they are qualitatively valid for smaller values of mass ratios as shown in Fig.~\ref{fig:kc_theta}, where we present a numerically calculated critical wave vector as a function of $r_s$.

Results shown in Figs.~\ref{fig:loss_critical} and~\ref{fig:kc_theta} imply a dramatic scenario where the anisotropic plasmon damping could be controlled either by varying the magnitude of the wave number (as for regular isotropic plasmons also, with $q>q_c$ being the Landau damping regime) or by varying the direction of the plasmon propagation at fixed wave number. Rotating the plasmon mode propagation direction could convert a totally damped plasmon into a totally undamped mode and vice versa. This spectacular new prediction could be easily verified experimentally in anisotropic electronic materials.
For example, one can use momentum-resolved inelastic electron energy-loss or inelastic light scattering or infrared optical spectroscopy (the precise technique depends on the relevant plasmon energy scale being probed) to obtain the loss function for a fixed momentum. The measured loss function should vary by several orders of magnitude as a function of plasmon propagation direction, as shown in the upper two figures in Fig.~\ref{fig:loss_critical}

Before concluding, we mention that we have also calculated the anisotropic plasmon dispersion in linearly dispersing massless Dirac-Weyl type 2D and 3D systems, where the electron energy dispersion is linear, rather than parabolic.  So, the anisotropy here is in the effective Fermi velocity $(v_x, v_y, v_z)$, not in any effective mass. The theory follows what we describe above for the parabolic systems, so we just quote the long-wavelength anisotropic plasmon dispersions for Dirac-like linearly dispersing bands:
\begin{equation}
    \omega_\mathrm{2D}(\bm q\rightarrow0)=\sqrt{r_s}
    \left(n\pi\right)^{\frac{1}{4}}
    \left| \bm  V  \hat{\bm n} \right|\sqrt{q},
\end{equation}
\begin{equation}
    \omega_\mathrm{3D}(\bm q\rightarrow0)=\sqrt{r_s}
    \left(\frac{32\pi}{3}\right)^{\frac{1}{6}}n^{1/3}
    \left| \bm  V  \hat{\bm n} \right|.
\end{equation}
Here $r_s=e^2/\hbar v_\mathrm{DOS}$ is the effective fine structure constant with $v_\mathrm{DOS}=(v_xv_y)^{1/2}$ in 2D and $(v_x v_y v_z)^{1/3}$ in 3D, and $\hat{\bm n}$ is the unit vector defining the plasmon propagation direction. $\bm  V$ is the velocity matrix defined to be $V_{ij}=v_i\delta_{ij}$ where $i=x,y,z$. In the isotropic limit of $v_x=v_y=v_z=v$, this reduces to the well-known Dirac plasmon dispersion, which depends on $\hbar$ (through the $r_s$ dependence) even at long wavelength since Dirac systems are quintessentially quantum mechanical with no classical analogs \cite{Sarma2009}.

To conclude, we have developed the theory for anisotropic plasmons in electronic materials by explicitly taking into account the anisotropic effective mass (or Fermi velocity) dispersion of the band energy.  We find both quantitative and qualitative effects arising from anisotropy, which cannot be captured in an isotropic approximation. In particular, the Landau damping depends crucially on the plasmon propagation direction, with the manifest possibility of a dramatic effect where the plasmon, at the same wavenumber, may be totally undamped in one direction and totally damped in another direction. First principles numerical band calculations for plasmon dispersion are unable to capture the new physics we predict. Our work should be of relevance to anisotropic metals and semiconductors as well as to new 2D materials (e.g., black-phosphorus), which are strongly anisotropic. 

\acknowledgments
This work is supported by the Laboratory for Physical Sciences.

\bibliographystyle{apsrev4-1}
\bibliography{ref}

\begin{thebibliography}{25}%
\makeatletter
\providecommand \@ifxundefined [1]{%
 \@ifx{#1\undefined}
}%
\providecommand \@ifnum [1]{%
 \ifnum #1\expandafter \@firstoftwo
 \else \expandafter \@secondoftwo
 \fi
}%
\providecommand \@ifx [1]{%
 \ifx #1\expandafter \@firstoftwo
 \else \expandafter \@secondoftwo
 \fi
}%
\providecommand \natexlab [1]{#1}%
\providecommand \enquote  [1]{``#1''}%
\providecommand \bibnamefont  [1]{#1}%
\providecommand \bibfnamefont [1]{#1}%
\providecommand \citenamefont [1]{#1}%
\providecommand \href@noop [0]{\@secondoftwo}%
\providecommand \href [0]{\begingroup \@sanitize@url \@href}%
\providecommand \@href[1]{\@@startlink{#1}\@@href}%
\providecommand \@@href[1]{\endgroup#1\@@endlink}%
\providecommand \@sanitize@url [0]{\catcode `\\12\catcode `\$12\catcode
  `\&12\catcode `\#12\catcode `\^12\catcode `\_12\catcode `\%12\relax}%
\providecommand \@@startlink[1]{}%
\providecommand \@@endlink[0]{}%
\providecommand \url  [0]{\begingroup\@sanitize@url \@url }%
\providecommand \@url [1]{\endgroup\@href {#1}{\urlprefix }}%
\providecommand \urlprefix  [0]{URL }%
\providecommand \Eprint [0]{\href }%
\providecommand \doibase [0]{http://dx.doi.org/}%
\providecommand \selectlanguage [0]{\@gobble}%
\providecommand \bibinfo  [0]{\@secondoftwo}%
\providecommand \bibfield  [0]{\@secondoftwo}%
\providecommand \translation [1]{[#1]}%
\providecommand \BibitemOpen [0]{}%
\providecommand \bibitemStop [0]{}%
\providecommand \bibitemNoStop [0]{.\EOS\space}%
\providecommand \EOS [0]{\spacefactor3000\relax}%
\providecommand \BibitemShut  [1]{\csname bibitem#1\endcsname}%
\let\auto@bib@innerbib\@empty
\bibitem [{\citenamefont {Maier}(2007)}]{Maier2007}%
  \BibitemOpen
  \bibfield  {author} {\bibinfo {author} {\bibfnamefont {S.~A.}\ \bibnamefont
  {Maier}},\ }\href {\doibase 10.1007/0-387-37825-1} {\emph {\bibinfo {title}
  {{Plasmonics: Fundamentals and Applications}}}}\ (\bibinfo  {publisher}
  {Springer US},\ \bibinfo {address} {New York, NY},\ \bibinfo {year}
  {2007})\BibitemShut {NoStop}%
\bibitem [{\citenamefont {Pines}\ and\ \citenamefont {Bohm}(1952)}]{Pines1952}%
  \BibitemOpen
  \bibfield  {author} {\bibinfo {author} {\bibfnamefont {D.}~\bibnamefont
  {Pines}}\ and\ \bibinfo {author} {\bibfnamefont {D.}~\bibnamefont {Bohm}},\
  }\href {\doibase 10.1103/PhysRev.85.338} {\bibfield  {journal} {\bibinfo
  {journal} {Phys. Rev.}\ }\textbf {\bibinfo {volume} {85}},\ \bibinfo {pages}
  {338} (\bibinfo {year} {1952})}\BibitemShut {NoStop}%
\bibitem [{\citenamefont {Bohm}\ and\ \citenamefont {Pines}(1953)}]{Bohm1953}%
  \BibitemOpen
  \bibfield  {author} {\bibinfo {author} {\bibfnamefont {D.}~\bibnamefont
  {Bohm}}\ and\ \bibinfo {author} {\bibfnamefont {D.}~\bibnamefont {Pines}},\
  }\href {\doibase 10.1103/PhysRev.92.609} {\bibfield  {journal} {\bibinfo
  {journal} {Phys. Rev.}\ }\textbf {\bibinfo {volume} {92}},\ \bibinfo {pages}
  {609} (\bibinfo {year} {1953})}\BibitemShut {NoStop}%
\bibitem [{\citenamefont {Pines}\ and\ \citenamefont
  {Nozieres}(1966)}]{Pines1966}%
  \BibitemOpen
  \bibfield  {author} {\bibinfo {author} {\bibfnamefont {D.}~\bibnamefont
  {Pines}}\ and\ \bibinfo {author} {\bibfnamefont {P.}~\bibnamefont
  {Nozieres}},\ }\href@noop {} {\emph {\bibinfo {title} {{Quantum Liquids, Vol.
  1}}}}\ (\bibinfo  {publisher} {WA Benjamin, New York},\ \bibinfo {year}
  {1966})\BibitemShut {NoStop}%
\bibitem [{\citenamefont {Platzman}\ and\ \citenamefont
  {Wolff}(1973)}]{Platzman1973}%
  \BibitemOpen
  \bibfield  {author} {\bibinfo {author} {\bibfnamefont {P.~M.}\ \bibnamefont
  {Platzman}}\ and\ \bibinfo {author} {\bibfnamefont {P.~A.}\ \bibnamefont
  {Wolff}},\ }\href@noop {} {\emph {\bibinfo {title} {{Waves and interactions
  in solid state plasmas}}}},\ Vol.~\bibinfo {volume} {13}\ (\bibinfo
  {publisher} {Academic Press New York},\ \bibinfo {year} {1973})\BibitemShut
  {NoStop}%
\bibitem [{\citenamefont {Abrikosov}(1988)}]{Abrikosov1988}%
  \BibitemOpen
  \bibfield  {author} {\bibinfo {author} {\bibfnamefont {A.~A.}\ \bibnamefont
  {Abrikosov}},\ }\href@noop {} {\emph {\bibinfo {title} {{Fundamentals of the
  Theory of Metals}}}}\ (\bibinfo  {publisher} {North-Holland, Amsterdam},\
  \bibinfo {year} {1988})\BibitemShut {NoStop}%
\bibitem [{\citenamefont {Quinn}\ and\ \citenamefont {Yi}(2009)}]{Quinn2009}%
  \BibitemOpen
  \bibfield  {author} {\bibinfo {author} {\bibfnamefont {J.~J.}\ \bibnamefont
  {Quinn}}\ and\ \bibinfo {author} {\bibfnamefont {K.-S.}\ \bibnamefont {Yi}},\
  }\href {\doibase 10.1007/978-3-540-92231-5} {\emph {\bibinfo {title} {Solid
  State Physics. Principles and Modern Applications}}}\ (\bibinfo  {publisher}
  {Springer-Verlag, Berlin},\ \bibinfo {year} {2009})\BibitemShut {NoStop}%
\bibitem [{\citenamefont {Abrikosov}\ \emph {et~al.}(2012)\citenamefont
  {Abrikosov}, \citenamefont {Gorkov},\ and\ \citenamefont
  {Dzyaloshinski}}]{Abrikosov2012}%
  \BibitemOpen
  \bibfield  {author} {\bibinfo {author} {\bibfnamefont {A.~A.}\ \bibnamefont
  {Abrikosov}}, \bibinfo {author} {\bibfnamefont {L.~P.}\ \bibnamefont
  {Gorkov}}, \ and\ \bibinfo {author} {\bibfnamefont {I.~E.}\ \bibnamefont
  {Dzyaloshinski}},\ }\href@noop {} {\emph {\bibinfo {title} {{Methods of
  quantum field theory in statistical physics}}}}\ (\bibinfo  {publisher}
  {Courier Corporation},\ \bibinfo {year} {2012})\BibitemShut {NoStop}%
\bibitem [{\citenamefont {Fetter}\ and\ \citenamefont
  {Walecka}(2012)}]{Fetter2012}%
  \BibitemOpen
  \bibfield  {author} {\bibinfo {author} {\bibfnamefont {A.~L.}\ \bibnamefont
  {Fetter}}\ and\ \bibinfo {author} {\bibfnamefont {J.~D.}\ \bibnamefont
  {Walecka}},\ }\href@noop {} {\emph {\bibinfo {title} {{Quantum theory of
  many-particle systems}}}}\ (\bibinfo  {publisher} {Courier Corporation},\
  \bibinfo {year} {2012})\BibitemShut {NoStop}%
\bibitem [{\citenamefont {Mahan}(2000)}]{Mahan2013}%
  \BibitemOpen
  \bibfield  {author} {\bibinfo {author} {\bibfnamefont {G.~D.}\ \bibnamefont
  {Mahan}},\ }\href {\doibase 10.1007/978-1-4757-5714-9} {\emph {\bibinfo
  {title} {{Many-Particle Physics}}}}\ (\bibinfo  {publisher} {Springer US},\
  \bibinfo {address} {Boston, MA},\ \bibinfo {year} {2000})\BibitemShut
  {NoStop}%
\bibitem [{\citenamefont {Pines}(2018)}]{Pines2018}%
  \BibitemOpen
  \bibfield  {author} {\bibinfo {author} {\bibfnamefont {D.}~\bibnamefont
  {Pines}},\ }\href {\doibase 10.1201/9780429500855} {\emph {\bibinfo {title}
  {{Elementary Excitations In Solids}}}}\ (\bibinfo  {publisher} {CRC Press},\
  \bibinfo {year} {2018})\BibitemShut {NoStop}%
\bibitem [{\citenamefont {Lindhard}(1954)}]{Lindhard1954}%
  \BibitemOpen
  \bibfield  {author} {\bibinfo {author} {\bibfnamefont {J.}~\bibnamefont
  {Lindhard}},\ }\href@noop {} {\bibfield  {journal} {\bibinfo  {journal} {Kgl.
  Danske Videnskab. Selskab Mat.-Fys.Medd.}\ }\textbf {\bibinfo {volume}
  {28}},\ \bibinfo {pages} {8} (\bibinfo {year} {1954})}\BibitemShut {NoStop}%
\bibitem [{\citenamefont {Nozi{\`{e}}res}\ and\ \citenamefont
  {Pines}(1958{\natexlab{a}})}]{Nozieres1958}%
  \BibitemOpen
  \bibfield  {author} {\bibinfo {author} {\bibfnamefont {P.}~\bibnamefont
  {Nozi{\`{e}}res}}\ and\ \bibinfo {author} {\bibfnamefont {D.}~\bibnamefont
  {Pines}},\ }\href {\doibase 10.1103/PhysRev.109.741} {\bibfield  {journal}
  {\bibinfo  {journal} {Phys. Rev.}\ }\textbf {\bibinfo {volume} {109}},\
  \bibinfo {pages} {741} (\bibinfo {year} {1958}{\natexlab{a}})}\BibitemShut
  {NoStop}%
\bibitem [{\citenamefont {Nozi{\`{e}}res}\ and\ \citenamefont
  {Pines}(1958{\natexlab{b}})}]{Nozieres1958a}%
  \BibitemOpen
  \bibfield  {author} {\bibinfo {author} {\bibfnamefont {P.}~\bibnamefont
  {Nozi{\`{e}}res}}\ and\ \bibinfo {author} {\bibfnamefont {D.}~\bibnamefont
  {Pines}},\ }\href {\doibase 10.1103/PhysRev.109.1062} {\bibfield  {journal}
  {\bibinfo  {journal} {Phys. Rev.}\ }\textbf {\bibinfo {volume} {109}},\
  \bibinfo {pages} {1062} (\bibinfo {year} {1958}{\natexlab{b}})}\BibitemShut
  {NoStop}%
\bibitem [{\citenamefont {Nozi{\`{e}}res}\ and\ \citenamefont
  {Pines}(1958{\natexlab{c}})}]{Nozieres1958b}%
  \BibitemOpen
  \bibfield  {author} {\bibinfo {author} {\bibfnamefont {P.}~\bibnamefont
  {Nozi{\`{e}}res}}\ and\ \bibinfo {author} {\bibfnamefont {D.}~\bibnamefont
  {Pines}},\ }\href {\doibase 10.1103/PhysRev.109.762} {\bibfield  {journal}
  {\bibinfo  {journal} {Phys. Rev.}\ }\textbf {\bibinfo {volume} {109}},\
  \bibinfo {pages} {762} (\bibinfo {year} {1958}{\natexlab{c}})}\BibitemShut
  {NoStop}%
\bibitem [{\citenamefont {Ehrenreich}\ and\ \citenamefont
  {Cohen}(1959)}]{Ehrenreich1959}%
  \BibitemOpen
  \bibfield  {author} {\bibinfo {author} {\bibfnamefont {H.}~\bibnamefont
  {Ehrenreich}}\ and\ \bibinfo {author} {\bibfnamefont {M.~H.}\ \bibnamefont
  {Cohen}},\ }\href {\doibase 10.1103/PhysRev.115.786} {\bibfield  {journal}
  {\bibinfo  {journal} {Phys. Rev.}\ }\textbf {\bibinfo {volume} {115}},\
  \bibinfo {pages} {786} (\bibinfo {year} {1959})}\BibitemShut {NoStop}%
\bibitem [{\citenamefont {Singwi}\ \emph {et~al.}(1968)\citenamefont {Singwi},
  \citenamefont {Tosi}, \citenamefont {Land},\ and\ \citenamefont
  {Sj{\"{o}}lander}}]{Singwi1968}%
  \BibitemOpen
  \bibfield  {author} {\bibinfo {author} {\bibfnamefont {K.~S.}\ \bibnamefont
  {Singwi}}, \bibinfo {author} {\bibfnamefont {M.~P.}\ \bibnamefont {Tosi}},
  \bibinfo {author} {\bibfnamefont {R.~H.}\ \bibnamefont {Land}}, \ and\
  \bibinfo {author} {\bibfnamefont {A.}~\bibnamefont {Sj{\"{o}}lander}},\
  }\href {\doibase 10.1103/PhysRev.176.589} {\bibfield  {journal} {\bibinfo
  {journal} {Phys. Rev.}\ }\textbf {\bibinfo {volume} {176}},\ \bibinfo {pages}
  {589} (\bibinfo {year} {1968})}\BibitemShut {NoStop}%
\bibitem [{\citenamefont {Sawada}\ \emph {et~al.}(1957)\citenamefont {Sawada},
  \citenamefont {Brueckner}, \citenamefont {Fukuda},\ and\ \citenamefont
  {Brout}}]{Sawada1957}%
  \BibitemOpen
  \bibfield  {author} {\bibinfo {author} {\bibfnamefont {K.}~\bibnamefont
  {Sawada}}, \bibinfo {author} {\bibfnamefont {K.~A.}\ \bibnamefont
  {Brueckner}}, \bibinfo {author} {\bibfnamefont {N.}~\bibnamefont {Fukuda}}, \
  and\ \bibinfo {author} {\bibfnamefont {R.}~\bibnamefont {Brout}},\ }\href
  {\doibase 10.1103/PhysRev.108.507} {\bibfield  {journal} {\bibinfo  {journal}
  {Phys. Rev.}\ }\textbf {\bibinfo {volume} {108}},\ \bibinfo {pages} {507}
  (\bibinfo {year} {1957})}\BibitemShut {NoStop}%
\bibitem [{\citenamefont {DuBois}(1959{\natexlab{a}})}]{DuBois1959a}%
  \BibitemOpen
  \bibfield  {author} {\bibinfo {author} {\bibfnamefont {D.}~\bibnamefont
  {DuBois}},\ }\href {\doibase 10.1016/0003-4916(59)90016-8} {\bibfield
  {journal} {\bibinfo  {journal} {Ann. Phys. (N. Y).}\ }\textbf {\bibinfo
  {volume} {7}},\ \bibinfo {pages} {174} (\bibinfo {year}
  {1959}{\natexlab{a}})}\BibitemShut {NoStop}%
\bibitem [{\citenamefont {DuBois}(1959{\natexlab{b}})}]{DuBois1959}%
  \BibitemOpen
  \bibfield  {author} {\bibinfo {author} {\bibfnamefont {D.}~\bibnamefont
  {DuBois}},\ }\href {\doibase 10.1016/0003-4916(59)90062-4} {\bibfield
  {journal} {\bibinfo  {journal} {Ann. Phys. (N. Y).}\ }\textbf {\bibinfo
  {volume} {8}},\ \bibinfo {pages} {24} (\bibinfo {year}
  {1959}{\natexlab{b}})}\BibitemShut {NoStop}%
\bibitem [{\citenamefont {Hwang}\ and\ \citenamefont {{Das
  Sarma}}(2007)}]{Hwang2007}%
  \BibitemOpen
  \bibfield  {author} {\bibinfo {author} {\bibfnamefont {E.~H.}\ \bibnamefont
  {Hwang}}\ and\ \bibinfo {author} {\bibfnamefont {S.}~\bibnamefont {{Das
  Sarma}}},\ }\href {\doibase 10.1103/PhysRevB.75.205418} {\bibfield  {journal}
  {\bibinfo  {journal} {Phys. Rev. B}\ }\textbf {\bibinfo {volume} {75}},\
  \bibinfo {pages} {205418} (\bibinfo {year} {2007})}\BibitemShut {NoStop}%
\bibitem [{\citenamefont {{Das Sarma}}\ and\ \citenamefont
  {Hwang}(2009)}]{Sarma2009}%
  \BibitemOpen
  \bibfield  {author} {\bibinfo {author} {\bibfnamefont {S.}~\bibnamefont {{Das
  Sarma}}}\ and\ \bibinfo {author} {\bibfnamefont {E.~H.}\ \bibnamefont
  {Hwang}},\ }\href {\doibase 10.1103/PhysRevLett.102.206412} {\bibfield
  {journal} {\bibinfo  {journal} {Phys. Rev. Lett.}\ }\textbf {\bibinfo
  {volume} {102}},\ \bibinfo {pages} {206412} (\bibinfo {year}
  {2009})}\BibitemShut {NoStop}%
\bibitem [{\citenamefont {Yu}\ \emph {et~al.}(2018)\citenamefont {Yu},
  \citenamefont {Guo}, \citenamefont {Xia},\ and\ \citenamefont {{Garc{\'{i}}a
  de Abajo}}}]{Yu2018}%
  \BibitemOpen
  \bibfield  {author} {\bibinfo {author} {\bibfnamefont {R.}~\bibnamefont
  {Yu}}, \bibinfo {author} {\bibfnamefont {Q.}~\bibnamefont {Guo}}, \bibinfo
  {author} {\bibfnamefont {F.}~\bibnamefont {Xia}}, \ and\ \bibinfo {author}
  {\bibfnamefont {F.~J.}\ \bibnamefont {{Garc{\'{i}}a de Abajo}}},\ }\href
  {\doibase 10.1103/PhysRevLett.121.057404} {\bibfield  {journal} {\bibinfo
  {journal} {Phys. Rev. Lett.}\ }\textbf {\bibinfo {volume} {121}},\ \bibinfo
  {pages} {057404} (\bibinfo {year} {2018})}\BibitemShut {NoStop}%
\bibitem [{\citenamefont {Brey}\ \emph {et~al.}(2020)\citenamefont {Brey},
  \citenamefont {Stauber}, \citenamefont {Mart{\'{i}}n-Moreno},\ and\
  \citenamefont {G{\'{o}}mez-Santos}}]{Brey2020}%
  \BibitemOpen
  \bibfield  {author} {\bibinfo {author} {\bibfnamefont {L.}~\bibnamefont
  {Brey}}, \bibinfo {author} {\bibfnamefont {T.}~\bibnamefont {Stauber}},
  \bibinfo {author} {\bibfnamefont {L.}~\bibnamefont {Mart{\'{i}}n-Moreno}}, \
  and\ \bibinfo {author} {\bibfnamefont {G.}~\bibnamefont
  {G{\'{o}}mez-Santos}},\ }\href {\doibase 10.1103/PhysRevLett.124.257401}
  {\bibfield  {journal} {\bibinfo  {journal} {Phys. Rev. Lett.}\ }\textbf
  {\bibinfo {volume} {124}},\ \bibinfo {pages} {257401} (\bibinfo {year}
  {2020})}\BibitemShut {NoStop}%
\bibitem [{\citenamefont {Stern}(1967)}]{Stern1967}%
  \BibitemOpen
  \bibfield  {author} {\bibinfo {author} {\bibfnamefont {F.}~\bibnamefont
  {Stern}},\ }\href {\doibase 10.1103/PhysRevLett.18.546} {\bibfield  {journal}
  {\bibinfo  {journal} {Phys. Rev. Lett.}\ }\textbf {\bibinfo {volume} {18}},\
  \bibinfo {pages} {546} (\bibinfo {year} {1967})}\BibitemShut {NoStop}%
\end{thebibliography}%

\end{document}